# Theoretical investigation of the feasibility of electronic mechanism for superconducting pairing in overdoped cuprates with Fermi liquid like and Marginal Fermi liquid- like descriptions for the normal phase

Soumi Roy Chowdhury and Ranjan Chaudhury

Department of Condensed Matter Physics and Material Sciences, S N Bose National Centre for Basic Sciences, Saltlake, Sector-III, Block-JD, Kolkata 700106, India

**Abstract:**

Cooper's formalism for fermionic pairing has been revisited considering upto $3^{rd}$ neighbour hopping terms, firstly with a Fermi liquid like background on a square lattice keeping in mind the overdoped cuprates. Then the whole scheme is repeated with a Marginal Fermi liquid- like background, taking into account the self energy correction of the mobile electrons to include a more realistic density of states in the calculation. Detailed comparison of our theoretical results with those from experiments strongly supports the Marginal Fermi liquid- like character of the normal phase with exciton mediated superconducting pairing in the concerned materials, in the lightly overdoped phase.

**Introduction:**

Theoretical demonstrations of superconductivity in overdoped cuprates supported by Fermi liquid (FL) like description of parental normal phase, were presented by many researchers till now [1-6]. Cuprates undergo transitions between different phases- from insulating with long range antiferromagnetism to superconductor with normal phase of non-Fermi liquid nature and then again to superconductor having Fermi liquid (FL) characterized normal phase, with increase in doping percentage and lowering of temperature. It has been argued that with increasing concentration of mobile holes the fermionic correlation in the system becomes weaker compared to the band width and the cuprate system becomes a better metal [8]. By measuring the transport of both heat and charge in the normal state at very low temperature, experimentalists were able to verify that one of the hole doped cuprates in the over doped regime obeys the Wiedemann-

Franz (WF) law, which is quite a definitive signature of FL theory [1]. The specific heat and magnetic susceptibility at the ideal composition are proportional and constant respectively with respect to temperature, consistent with Fermi liquid [9].

One prominent class of examples with deviation from the Fermi liquid description is the strange metal phase of the cuprate superconductors, which refers to the metallic state above the superconducting transition temperature Tc in the vicinity of optimal doping. The strange metal phase exhibits thermodynamic and transport behavior significantly different from those of an ordinary metal. A particularly striking property of the strange metal phase is that the electrical resistivity increases linearly with temperature, in contrast to the quadratic temperature dependence of an ordinary metal. This remarkably simple behavior is existing over a wide range of temperature, appearing in all cuprate superconductors. The anomalous behavior can be answered with the help of photoemission experiments which can probe directly a Fermi surface and its low energy excitations. In an Angular Resolved Photoemission Spectropy (ARPES) experiment, incident photons knock out electrons/ holes from the sample and the intensity $I(\omega, k)$ of the electron/ hole beam is proportional to $A(\omega, k)f(\omega_k)$, where $f(\omega_k)$ is the Fermi-Dirac distribution and $A(\omega, k)$ is the electron spectral function defined by

$$A(\omega, k) = \left(-\frac{1}{\pi}\right) \text{Im } G(\omega, k)$$

ARPES experiments indicate that a Fermi surface still exists, but excitations exhibit a much broader peak than that for a Fermi liquid [10]. The experimental results can be fitted well to the following expression, postulated as "Marginal Fermi liquid" (MFL) in,

$$G(\omega, k) = h/\{\omega - v_F(k - k_F) + \Sigma(\omega, k)\}$$

with the self-energy $\Sigma(\omega, T)$ of the fermionic carriers behaves for $\omega > T$ like $\Sigma^{Re}(\omega, T) \sim \omega \ln|\omega|$ and $\Sigma^{Im}(\omega, T) \sim |\omega|$ in contrast to ordinary Fermi liquid theory where $\Sigma^{Re}(\omega, T) \sim \omega$ and $\Sigma^{Im}(\omega, T) \sim \omega^2$ [11]. We see from this that the system appears to possess gapless excitations of dispersion relation $\omega_{ex} = v_F(k - k_F)$ ). However, the decay rate $\tau$ of such excitations, which is given by the imaginary part of $\Sigma$, is now proportional to $\omega$ in contrast to $\omega^2$ for a Fermi liquid. The decay rate, which is comparable to $\omega$, is so large, that an excitation will already have decayed before it can propagate far enough (i.e. one wavelength) to show its particle-like properties. As a result, such an excitation can no longer be treated as a quasiparticle.

An electron/hole in a medium like highly correlated system is dressed by a cloud of excitations and acquires a different effective mass, but still behaves as a single particle excitation or quasiparticle. These interacting effects with the other excitations are manifested through the self energy of a single particle. Now it is worth emphasizing that the exact calculation of self energy is an extremely difficult task. By causality, the real and imaginary parts of self-energy are related by Kramers-Kronig dispersion relations. In principle, if the full spectral function A(k, ω) is known, one could perform an inversion to obtain the full self energy spectra by using the Kramers-Kronig transformation [12] . However, such transformation has limitations and requires the spectrum from −∞ to +∞ in energy. Unfortunately, clean ARPES data from doped cuprates can usually be obtained from Fermi level to around half of the band width where complication of valance bands will come in. So the process requires further to take a cut-off/extension model at energies above the existing data points.

We have done our calculation for checking the probable pairing between two fermions in both FL- like and MFL- like backgrounds with phononic mechanism as well as excitonic mechanism. We extract various important physical quantities for cuprate superconductors, such as coherence length, coupling constant and critical temperature. After a careful quantitative comparison of our theoretical results with experimental ones, we arrive at the conclusion of the MFL like normal phase in the lightly overdoped cuprates. Theoretical demonstrations justifying the substantial s- wave component and experiments with Raman scattering and tunneling spectra back the conception of a sturdy existence of s- wave symmetry along with other pairing symmetry [13]. We have examined the adequacy and feasibility of the s- wave symmetry in intra- layer pair formation though the presence of a d-wave symmetry is not ruled out.

**Mathematical Formalism:** The analogous form of pairing equation including the highly asymmetrical energy dispersion relation in two dimensional tight binding hamiltonian by Cooper's treatment reads [14]:

$$\left(\frac{u}{A}\right) = U = \frac{1}{\int_{\epsilon_F}^{\epsilon_F + 8t(1-\delta) + (\hbar\omega_{boson} - 8t(1-\delta)\theta(8t(1-\delta) - \hbar\omega_{boson})} \frac{D(\widetilde{\epsilon}_k)\, d\widetilde{\epsilon}_k}{(2\widetilde{\epsilon}_k - 2\epsilon_F + |W|)}} \qquad (1)$$

Generalised 2D energy dispersion can be expressed in tight binding representation as:

$$\epsilon_k = \epsilon_0 - 2t(\cos k_x a + \cos k_y a) + 4t'(\cos k_x a \cdot \cos k_y a) - 2t''(\cos 2k_x a + \cos 2k_x a) \quad (2)$$

$$\tilde{\epsilon}_k = \epsilon_k - \epsilon_F$$

where t, t' and t'' are the single electron hopping parameters corresponding to the nearest neighbour, next nearest neighbour and the 3$^{rd}$ neighbour respectively on a square lattice. The quantity $(-u\delta(\vec{r}_1 - \vec{r}_2))$ with u>0, is the attractive contact interaction whose fourier transform being equal to (– u/A) (denoted by operating U) within the small energy transfer region $\hbar\omega_{boson}$ (the characteristic energy of the boson mediating attraction) above the Fermi surface where the pairing would take place [15]; D($\omega$) is the density of states; r is the area of the 2D system in consideration (A $\to \infty$ for a macroscopic system); |W| is the pairing energy for the two mates constituting a pair. $\epsilon_F$ is the Fermi energy and  $\delta$ is the band filling factor. The equation (1) represents the pairing situation in the 'passive Fermi Sea' background. Now to examine the realistic concept of pair formation in the presence of an active Fermi sea, we have to include the effect of Pauli's exclusion principle or rather Pauli blocking of the phase space. For this a factor $1 - f_{-k+q/2} - f_{k+q/2}$ has to be incorporated in the 2D single pairing Hamiltonian (equation 1b) where $f_{k+q/2}$ is the probability that there is a carrier of momentum $k + q/2$ above the Fermi level. Therefore the modified equation for pairing with finite centre of mass momentum $\hbar q$ becomes

$$1 = U \sum_{\vec{k}} \frac{1 - f_{-k+q/2} - f_{k+q/2}}{-E + 2\epsilon_k}$$

This leads to following equation→

$$1 = U \sum_{\vec{k}} \frac{-1 + \theta\left(\epsilon_{k+\frac{q}{2}} - \epsilon_k\right) + \theta(\epsilon_{-k+\frac{q}{2}} - \epsilon_k)}{(|W| + 2\tilde{\epsilon}_k)}$$

The characteristic of the Theta function determines the range of pairing and finally the pairing equation corresponding to this 'active Fermi sea' background becomes:

$$\left(\frac{u}{A}\right) = \left(\frac{\pi^2 B}{2}\right) \frac{1}{-\int_0^{\hbar\omega_{boson}} \frac{K\sqrt{1-\left(\frac{\widetilde{\epsilon}_k + \epsilon_F - \epsilon_0}{2t}\right)^2} d\widetilde{\epsilon}_k}{(|W| + 2\widetilde{\epsilon}_k)} + \int_{atqSin(ka)}^{\hbar\omega_{boson}} \frac{K\sqrt{1-\left(\frac{\widetilde{\epsilon}_k + \epsilon_F - \epsilon_0}{2t}\right)^2} d\widetilde{\epsilon}_k}{(|W| + 2\widetilde{\epsilon}_k)} + \int_{-atqSin(ka)}^{\hbar\omega_{boson}} \frac{K\sqrt{1-\left(\frac{\widetilde{\epsilon}_k + \epsilon_F - \epsilon_0}{2t}\right)^2} d\widetilde{\epsilon}_k}{(|W| + 2\widetilde{\epsilon}_k)}} \quad (3)$$

It may be remarked that our lattice Hamiltonian for Cooper pairing looks somewhat similar to negative - U Hubbard model, although the attraction here operates only within a finite energy interval. It may be remarked here that to start with we keep both the possibilities of bosonic mechanism viz. electronic and phononic open under s-wave pairing scheme.

For the MFL scenario

$$D(\omega) = \sum_k -\text{Im}[G(\omega, k)/\pi]/N^2 \quad (4)$$

Where $G(\omega, k) = [1/(\omega - \epsilon_k - \Sigma(k, \omega))]$ (5)

N is number of lattice sites. The above form of the Green's function for a single fermion on the 2D lattice, includes the self energy corrections $\Sigma(k, \omega)$ to implement the many body effects within the system. Researchers in general parameterize only ARPES data and suggest various models of self energy consistent with the fitted plots [16]. We choose the simple form of the single particle normal phase self energy as proposed by Varma and is characterized by the functional form of

$$\Sigma(k, \omega) = \Sigma^{Re}(k, \omega) + i\Sigma^{Im}(k, \omega) \quad (6)$$

$$\Sigma^{Re}(k, \omega) \propto \omega \ln\left(\frac{|\omega|}{\omega_c}\right) \text{ and } \Sigma^{Im}(k, \omega) \propto \omega \text{ for } |\omega| > T$$

We have slightly modified the real part and take it as

$$\Sigma^{Re}(k, \omega) = g\omega \ln((|\omega| + \varepsilon)/\omega_c) \text{ for } |\omega| > T, \quad (7)$$

$\varepsilon$ is a doping dependent term which increases with reduced wave vector. This is associated with the imaginary self energy which is derived by plugging Kramers- Cronig relation viz.

$$\Sigma^{Im}(k,\omega) = -\left(\frac{1}{\pi}\right)\int_{-\infty}^{\infty}[\frac{\Sigma^{Re}(k,\omega)d\omega}{(\omega-\omega')}] \qquad (8)$$

Instead of infinite extent the range of the integration is kept limited in between the band width though. We have explicitly checked the applicability of this expression by plotting the real part of self energy against frequency. This small change in the proposed MFL self energy formula and the associated parameterization gives an almost accurate matching for the ARPES spectra obtained from Bi2212 sample [17].

Fig 1: A small remodeling of the MFL self energy has been done on the basis of analysis of line shape spectra in Bi2212 sample.

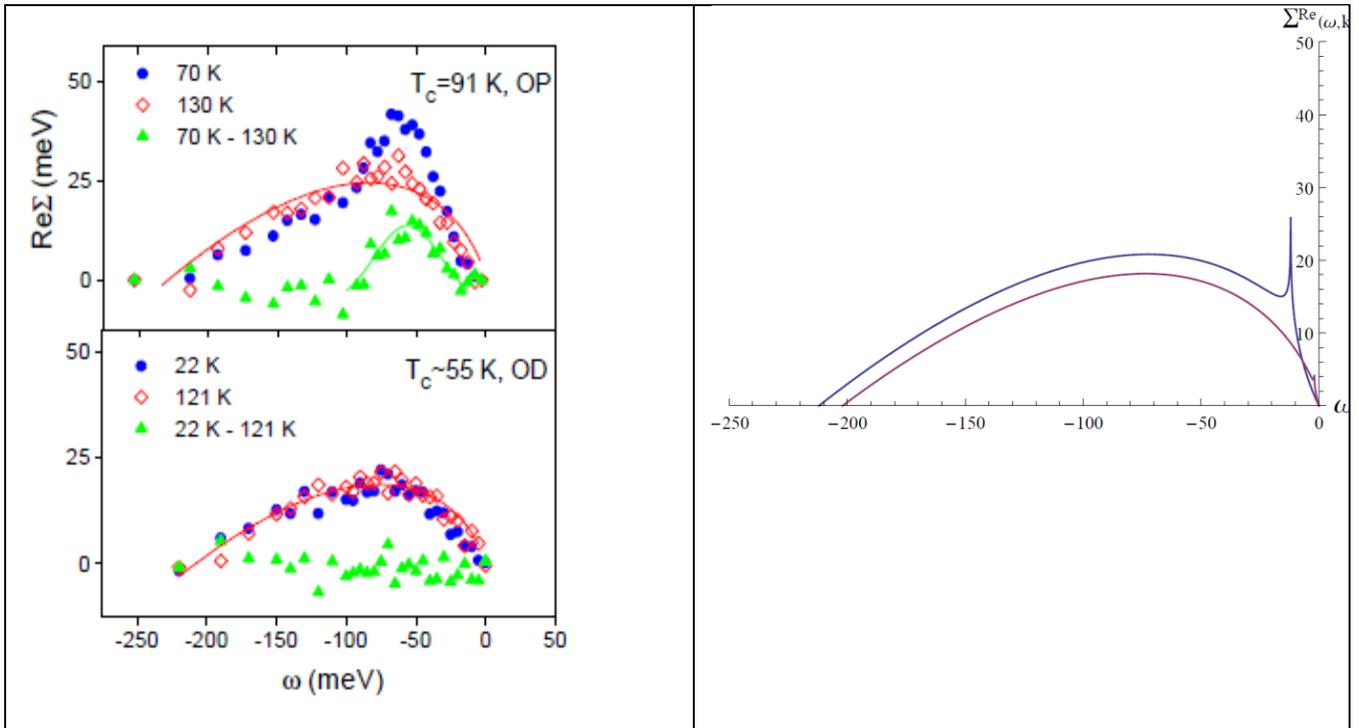

| Fig: 1a Real part of self energy as a function of energy for the superconducting (blue dots) and normal states (open red diamonds) for the optimally doped and overdoped samples, as obtained by reference indicated [8]. The solid lines through the normal state data represent MFL fits to the data. | Fig: 1b Remodelling of $\Sigma^{Re}$ using our formula for Bi2212 The red line (below one) is for overdoped and the blue line (above one) corresponds to optimally doped region. |
|---|---|

The corresponding imaginary part of self energy viz. $\Sigma^{Im}(\omega,k)$ is obtained by incorporating the form of real part of self energy (from fig 1b) in Kramers Kronig relation in a full band width and is parameterized with the following function of $\omega$, to be incorporated in the Green's function.

$$f(\omega, k) = \left(-2.8 \ln\left[\frac{Abs[\omega + 1]}{500}\right]\right) \quad (9)$$

Here are the two plots for this function.

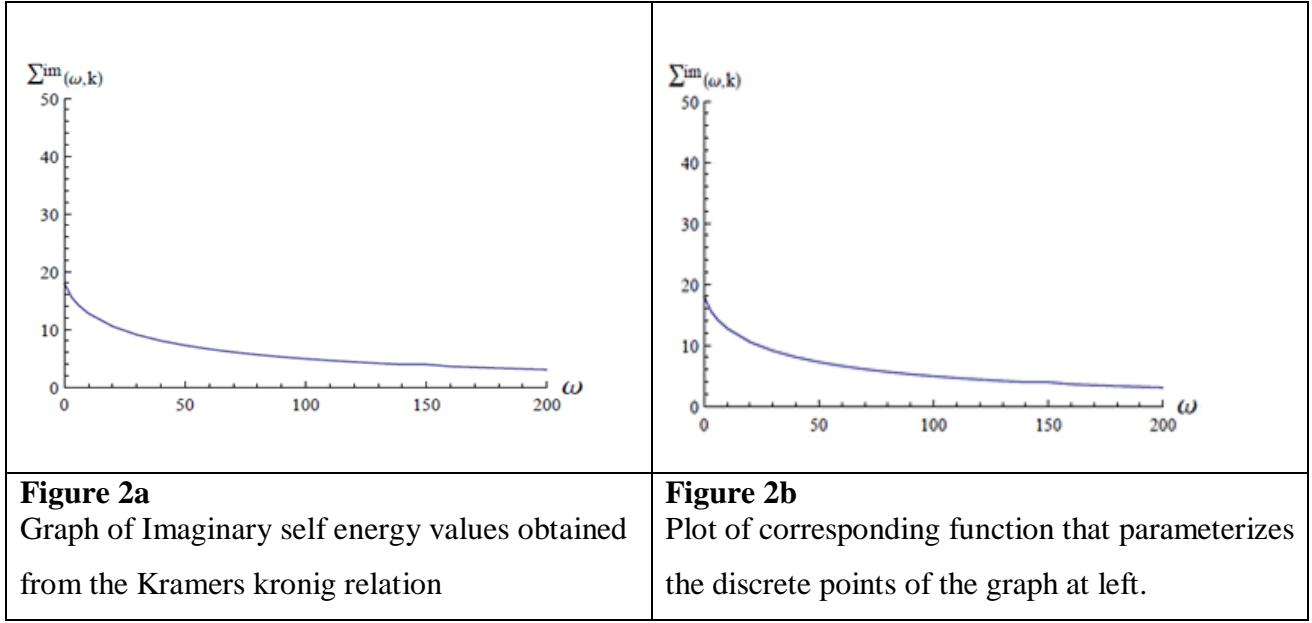

**Figure 2a**
Graph of Imaginary self energy values obtained from the Kramers kronig relation

**Figure 2b**
Plot of corresponding function that parameterizes the discrete points of the graph at left.

The fitting of imaginary self energy is quite tricky as it is considered to be linear with frequency but shows a visible deviation from linearity as obtained from different spectroscopic experimental result [9]. Our graph shows logarithmic nature of $\Sigma^{Im}(\omega,k)$ though, without any calculative manipulation, but behaves almost linearly over a big frequency region. A large number of experimentalists are involved in proper understanding of the normal state scattering rate, which governs the transport properties, although their approach has strong limitations. They measure or infer the scattering rate (proportional to $\Sigma^{Im}$) at a single energy ($\omega$) and do not extract its functional form. Other studies that investigate the scattering rate as a function of energy are limited to the nodal direction alone. The data presented in our paper provide a comprehensive measurement of the functional form of the scattering rate as a function of energy around the Fermi surface. More importantly, our proposed form for $\Sigma^{Im}(\omega,k)$ as given in equation (9) and Figure 2 show a striking similarity with those extracted from ARPES (Figs. 3a-3d)

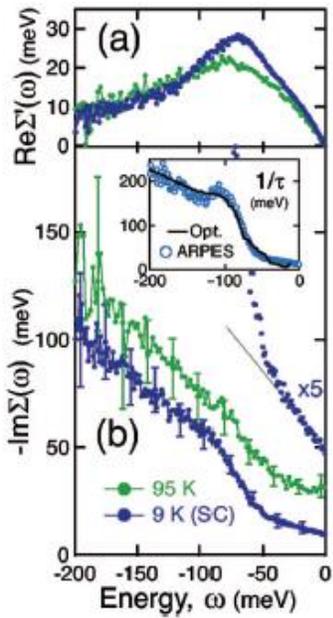

Figure 3a
Real part and Imaginary part of selfenergy as obtained by angle-resolved photoemission spectroscopy for overdoped cuprates [18].

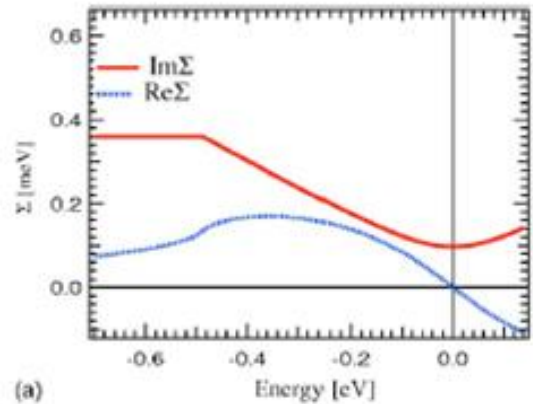

Figure 3b
The fitting of the EDC data for the energy dependence of the scattering rate of the high temperature cuprate superconductors obtained from angle resolved photoemission spectroscopy [19].

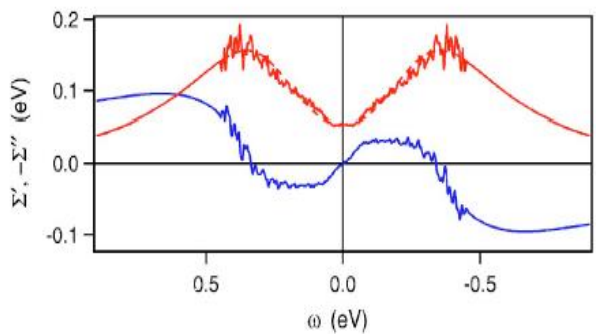

Figure 1c
The results of the fitting procedure for Bi(Pb)-2212 OD75: for real and imaginary parts of the self-energy [20].

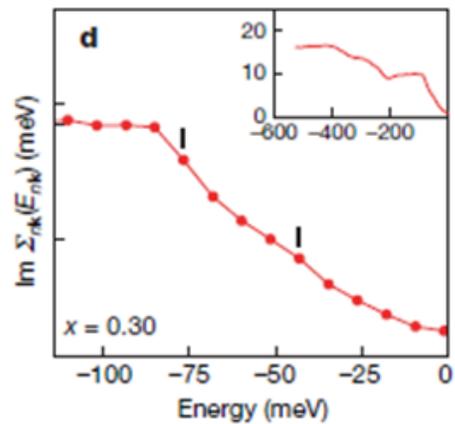

Figure 3d
Calculated imaginary part of self energy for overdoped LSCO from ARPES data.[21]

The density of states (DOS) associated with the total self energy of the following form is plotted and resemble the unusual dip that observed in photoemission experiments on Bi2212 [10]. Therefore the total self energy takes the form

$$\Sigma(k,\omega) = \Sigma^{real}(\omega,k) + i\,\Sigma^{im}(\omega,k) = -0.24\left(\omega\,\text{Ln}\left[\frac{\text{Abs}[\omega+12]}{200}\right]\right) - i\left(2.8\,\ln\left[\frac{\text{Abs}[\omega+1]}{500}\right]\right) \qquad (10)$$

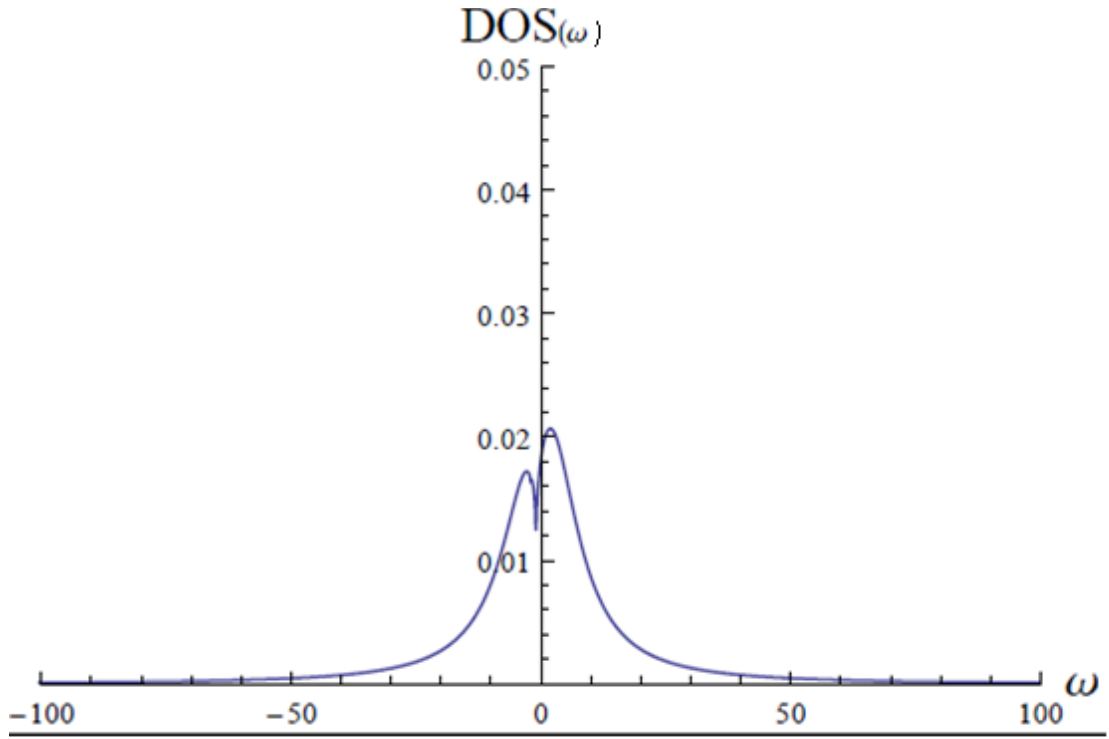

**Figure 3 Plot of density of states vs energy using our theoretical expression of self energy**

It must be kept in mind however, that the pairing equation is strictly not valid at $\omega = \epsilon_F$ as the quasi particle weight vanishes there [22]

In our formalism the band is completely empty (i.e filling factor δ=0) when doping concentration is 1. Then introduction of carrier raises the degree of band filling with δ=1 representing half filled band. So the lower portion and the very upper portion of the band represents extremely overdoped and extremely underdoped region respectively as per the phase diagram (ordering temperature vs doping concentration) of the cuprate superconductors [23]. The region in between these two represents optimally to moderately overdoped region where superconductivity is likely to occur. The attractive coupling constant ($\lambda$) has been calculated using the formula [14]

$$\lambda = \left(\frac{u}{A}\right) D(\epsilon_F) \qquad (11)$$

The idea of the coherence length comes out clearly by understanding the spatial nature of the pair wave function associated with the finite centre of mass momentum case. The maximum allowed pairing wave vector '$q_{max}$' (defined by $|W| = 0$ for $q=q_{max}$) gives an estimate of the coherence length ($'\xi'$), which is of the order of reciprocal of $q_{max}$. The finite centre of mass momentum analogue of the pairing equation is

$$1 = u \sum_k \frac{1}{\{(\epsilon_{k+\frac{q}{2}} + \epsilon_{-k+\frac{q}{2}}) - E\}} \qquad (12)$$

With $\epsilon_{k+q/2} + \epsilon_{-k+q/2} = 2\epsilon_k + 2t'(aq)^2 \sin(k_x a)\sin(k_y a)$

The upper and lower limit of $\epsilon_k$ has been modified into

$\epsilon_F + \hbar\omega_{boson} - t'(aq)^2 \sin(k_x a)\sin(k_y a)/2$ and $\epsilon_F + t'(aq)^2 \sin(k_x a)\sin(k_y a)/2$ (13)

**Calculations and results:**

Scanty experimental data leads to an ambiguity in the energy of the boson, mediating pairing. We get good and realistic result in the range of 0.1- 0. 2 eV which in general is catagorised under electron originated pairing mechanism [24]. It may be remarked that a vertex correction can become quite important in the first principle calculation for 'u' if the bosonic energy (specially relevant for excitonic boson) becomes comparable to the Fermi energy, which is not the case here [16].

Both phononic and excitonic energy value has been incorporated in the pairing equation as the bosonic energy mediating pairing to check the substantiality of these two .The whole calculation is done in such a manner so that the basic Fermi Liquid- like criterion (U< 4t) is maintained. The relevant parameters presented in Table-1 are quite realistic [25].

Table-1: Parameters corresponding to pairing mechanism

| Parameters | Electronic mechanism | Phononic mechanism |
|---|---|---|
| Bosonic energy ($\hbar\omega_{boson}$) | 0.3 ev | 0.05 ev |

**FL like scenario**:

In conventional 3D isotropic Bardeen- Cooper- Schrieffer (BCS) superconductors, an energy gap $\Delta_{sc}$ opens below Tc with s-wave symmetry and $2\Delta_{sc}$ is the minimum energy required to break a Cooper pair. The magnitude of $2\Delta_{sc}$ is of the order of 50- 60 mev for overdoped cuprates [26,27]. Keeping this in mind, throughout our calculation the maximum value of |W| has been kept below 0.05 ev. The result of our calculations, shown in Figure 1b, indicates that the coupling constant lies in the intermediate range for electronic mechanism i.e. $\lambda\sim0.5$. For the same range of |W| the coupling constant lies in the strong range, i.e. $\lambda\sim0.7$ in case of phononic mechanism. Figure1a. is another graph for electronic mechanism which shows the variation of coupling constant for a higher range of |W|, i.e. a higher coupling constant (here upto 0.65) can be achieved via this mechanism. But increasing of |W| up to this level is not possible for phononic mechanism as this violates the FL criterion.

The above results of ours are very close to those obtained by using the conventional Cooper equation corresponding to the isotropic 3D case, viz.

$$|W| = 2\hbar\omega_{boson} \exp\left(-\frac{2}{\lambda}\right) \qquad (14)$$

From the above equation (4) the value of 0.007 ev for |W| gives a coupling constant of 0.45 in case of electronic mechanism. Graphically the above magnitude of |W| corresponds to a coupling constant of 0.5 from our calculation. Furthermore, |W| of 0.007 ev corresponds to a coupling constant of 0.7 from our calculation and 0.75 from equation (12) for phononic mechanism. We however consider the usage of this equation (12) here to be very limited as this equation is appropriate for 3D isotropic system, whereas the experimental system is truly quasi-2D or rather layered and our calculation is for pure 2D lattice. Nevertheless, we still compare the estimates of the coupling constant from the two approaches. The maximum value of the pairing energy can be upto the order of the superconducting gap according to the conventional theory of

superconductivity [28]. The estimates for coherence length obtained in this case, is presented in the Table below

Table 2: The calculated values of coherence length with hopping term of 0.2 ev

| δ | Electronic mechanism Value of ξ (in unit of 'a') | Phononic mechanism Value of ξ (in unit of 'a') |
|---|---|---|
| 0.5 | 666.67 | - |
| 0.99 | 19.6 | 110 |
| 1.4 | 692 | 3330.4 |
| 1.5 | - | 5265.16 |

Generally real high- $T_c$ cuprates (both in the under and overdoped regions) have short in-plane coherence length [29]. Both experimental results and our theoretical estimates (shown in Table-1 and 2) point to the fact that the in- plane coherence lengths corresponding to excitonic mechanism are much shorter than those in the conventional phonon driven 3D superconductors. Even in 2D phononic mechanism produces a much longer coherence length (as shown in Table 2). Lowering of the magnitude of coherence length in this case needs a large enhancement in the phonon mediated attractive interaction. This would violates the FL scenario and make it inappropriate for the overdoped phase. Therefore, the excitonic mechanism is the most feasible one for intra- layer pairing in the overdoped regime.

**MFL- like scenario :**

Following the above result obtained from FL- like scenario, we now do our calculation in MFL like scenario in near half filled band only with excitonic mechanism based pairing.

Table-3: Parameters corresponding to pairing in Bi2212

| Bosonic energy | Parameters | Bi2212 |
|---|---|---|
| 0.1 eV | Coupling Constant | 0.44 |
| | Coherence length (in unit of lattice constant) | 12 a |
| | Temperature | 115K |

| 0.15 eV | Coupling Constant | 0.38 |
|---|---|---|
| | Coherence length (in unit of lattice constant) | 14.287 a |
| | Temperature | 115K |

Strikingly the values of these three definitive signatures of overdoped cuprates -1) Short in-plane coherence length 2) Strong coupling constant and 3) High critical temperature are very much consistent with the experimentally obtained data for the relevant materials [30].

The critical temperature is the temperature correspondence of |W| which was extracted from the conventional Cooper's and BCS equations corresponding to the isotropic 3D case [14].

$$|W| = 2\hbar\omega_{boson} \frac{(KT_c)^2}{(1.13)^2 \hbar\omega_{boson}} \qquad (15)$$

We however consider the usage of these equations (15) here to be very limited as they are appropriate for 3D isotropic system, whereas our calculation is done in 2D. Nevertheless, we still compare the estimates of the coupling constant from the two approaches. Analysis of the available experimental results from Angle Resolved Photoemission Spectroscopy (ARPES) and from polar angular magnetoresistance oscillations show the presence of a 3D coherent electronic behaviour in overdoped phases of some of the cuprate superconductors. Investigation of the oscillations shows that at certain symmetry points however, the Fermi surface exhibits properties characteristic of 2D systems[31]. This striking form of the Fermi surface topography, provides a natural explanation for a wide range of anisotropic properties both in the normal and superconducting states of this system.

The other two parameters ($\lambda$ and $\xi$) are derived numerically by using the pairing equation. The coupling constant lies in the range, i.e. $\lambda \sim 0.4$ for Bi2212 and coherence lengths are of the order of $10^{-1}$ Å conforming to experimental values of till date [30].

In Cooper's model in continuum or in BCS Theory '$\lambda$' is independent of $\hbar\omega_{boson}$. According to McMillan's equation (obtained by simplification of Elliashberg's equation) however $\lambda$ is inversely proportional to $\hbar\omega_{boson}$[32]. This is in qualitative agreement with the variation seen in Table 3.

**Conclusion and Discussion:**

Both of the above calculations (with FL like and MFL like descriptions for the normal phase) show the most realistic signature of pairing in moderately overdoped (intermediate band filling range) region. The coherence length is smallest and is of the order of experimental value in the vicinity of half filling. Otherwise either it is too long or the calculation becames non tractable. If we try to look into the practical significance of our result then we recall that as per the experimental phase diagram of cuprate superconductors the underdoped normal regime doesn't obey the FL theory. In our calculation we get a non- tractable regime in the upper phase of band above a particular filling for a fixed bosonic energy and hopping parameter with the passive and active Fermi sea. The extremely high doped regime in the experimental phase diagram doesn't gives very long coherence length in low filling (extremely overdoped regime) indicating an unrealistic situation for pairing.

But our calculations following a MFL- like description of the normal phase with boson mediated pairing, reproduces real physics regarding the coherence length, coupling constant and critical temperature of the cuprate superconductors in general agreement with more sophisticated many body treatments like Eliashberg scheme and with experiments on real superconductors more sensitively than by FL like description does. From this viewpoint, the MFL like scenario with electronic mechanism seems to be more feasible one for the pairing in cuprates [33].

**References:**

1) C. Proust, E. Boaknin, R. W. Hill, L. Taillefer and A. P. Mackenzie Phys. Rev. Lett. **89** 14 (2002)

2) G.Kastrinakis Physica C **317-319** 497 (1999); G.Kastrinakis Physica C **340** 119-132 (2000); A.Damascelli Review of Modern Physics **75** (2003)

3) N.E. Hussey J.Phys. Cond. Matter **20** 123201 (2008)

4) A. F. Bangura, P. M. C. Rourke, T. M. Benseman, M. Matusiak, J. R. Cooper, N. E. Hussey, and A. Carrington Phys Rev. B **82** 140501 ® (2010)

5) M. Gurvitch and A.T. Fiory Phys. Rev. Lett **59** 12 (1987)

6) R. B. Laughlin Phys. Rev. Lett. **112** 017004 (2014)

7) T. M Mishonov, J. Indekeu and E. S. Penev J. Phys. Cond. Matter **15** 4429–4456 (2003)

8) N.C. Yeh, C.T. Chen, C.C. Fu, P. Seneor, Z. Huang, C.U. Jung, J.Y. Kim, M.S. Park, H.J. Kim, S.I. Lee, K. Yoshida, S. Tajima, G. Hammerl, J. Mannhart Physica C Superconductivity **367** 1–4 [2002]

9) C. M. Varma Phys. Rev. B **73**, 155113 (2006)

10) T. Yoshida, X. J. Zhou, H. Yagi, D. H. Lu, K. Tanaka, A. Fujimori, Z. Hussain, Z.-X. Shen, T. Kakeshita, H. Eisaki, S. Uchida, Kouji Segawa, A. N. Lavrov, and Yoichi Ando, Physica B 351, 250 (2004)

11) C. M. Varma, P. B. Littlewood, S. Schmitt-Rink, E. Abrahams, and A. E. Ruckenstein

Phys. Rev. Lett. **63**, 1996 (1989); Erratum Phys. Rev. Lett. **64**, 497 (1990)

Andrea Damascelli, Zahid Hussain, and Zhi-Xun Shen Rev. Mod. Phys. **75**, 473 2003

M. Franz and Z. Tesanovic Phys. Rev. Lett. **87**, 257003 (2001); Erratum Phys. Rev. Lett. **88**, 109902 (2002); M. E. Simon and C. M. Varma Phys. Rev. Lett**. 89** 247003 (2000)

**12)** W. Meevasana, F. Baumberger, K. Tanaka, F. Schmitt, W. R. Dunkel, D. H. Lu, S.-K. Mo, H. Eisaki, and Z. X. Shen Phys. Rev. B **77**, 104506 (2008)

M.R. Norman, H. Ding, H. Fretwell, M. Randeria, and J. C. Campuzano, Phys. Rev. B. **60,** 7585 (1999)

**13)** C.C. Chen, M. L. Teague, Z.J. Feng, R.T.P. Wu, N.C. Yeh APS March Meeting **58** 1 (2013)

J A. Skinta, M.S Kim, T. R. Lemberger, T. Greibe, and M. Naito Phys. Rev. Lett. **88** 207005 (2002)

14) S. Roy Chowdhury and R. Chaudhury  arXiv: 1506.08373v3 (2016); S. Roy Chowdhury and R. Chaudhury  Physica B 465 60-65 (2015)

**15)** S. Pal J.Phys.Commun**1** 055029 (2017)

**16)**  A. Kaminski, H. M. Fretwell, M. R. Norman, M. Randeria, S. Rosenkranz, U. Chatterjee, J. C. Campuzano, J. Mesot, T. Sato, T. Takahashi, T. Terashima, M. Takano, K. Kadowaki, Z. Z. Li, and H. Raffy Phys. Rev. B **71**, 014517 (2005)

17)  P.D. Johnson , T. Valla, A.V. Fedorov, Z. Yusof, B.O.Wells, Q. Li, A.R. Moodenbaugh, G.D Gu, N. Koshizuka, C. Kendziora, S. Jian, D.G.Hinks. Phys Rev Lett. 87 17:177007 (2001) 10) T. Yamasaki, K. Yamazaki, A. Ino, M. Arita, H. Namatame, M. Taniguchi, A. Fujimori, Z.-X. Shen, M. Ishikado, and S. Uchida Phys. Rev. B **75**, 140513(R) (2007)

18) T. Yamasaki, K. Yamazaki, A. Ino, M. Arita, H. Namatame, M. Taniguchi, A. Fujimori, Z.-X. Shen, M. Ishikado, and S. Uchida Phys. Rev. B **75**, 140513(R)  (2007)

19) A. Kaminski H. M. Fretwell M. R. Norman M. Randeria S. Rosenkranz U. Chatterjee J. C. Campuzano Physical Review B 71, 014517 (2005)

20) A.A.Kordyuk, S. V. Borisenko, A. Koitzsch, J. Fink, M. Knupfer and H. Berger  Physical Review B **71**, 214513 (2005)

21) F. Guistino,  M.L. Cohen, S.G. Louie, Arxiv: 0710.2146v2 (2007)

22) R. Chaudhury Can.J. Phys. **73** 497-504 (1995)

23) W. Meevasana, X. J. Zhou, S. Sahrakorpi, W. S. Lee, W.L. Yang, K. Tanaka, N. Mannella, T. Yoshida, D. H. Lu,Y. L. Chen, R. H. He, Hsin Lin, S. Komiya, Y. Ando, F.Zhou, W. X. Ti, J. W. Xiong, Z. X. Zhao, T. Sasagawa,T. Kakeshita, K. Fujita, S. Uchida, H. Eisaki, A. Fujimori,Z. Hussain, R. S. Markiewicz, A. Bansil, N. Nagaosa, J.Zaanen, T. P. Devereaux, and Z.-X. Shen, Phys. Rev. B**75**, 174506 (2007).

24) C. M. Varma Phys. Rev. B **73**, 155113 (2006)

25) A.S. Mishchenko and N.Nagaosa Phys Rev Lett **93** 3 (2004)

26) I. Zeljkovic, E. J. Main, T. L. Williams, M. C. Boyer, K. Chatterjee, W. D. Wise, Y. Yin, M. Zech, A. Pivonka, T. Kondo, T. Takeuchi, H. Ikuta, J. Wen, Z. Xu, G. D. Gu, E. W. Hudson and J. E. Hoffman Nature Materials **11** 585–589 (2012)

27) N. Miyakawa, P. Guptasarma, J. F. Zasadzinski, D. G. Hinks, and K. E. Gray Phys. Rev. Lett. **80**, 157 (1998)

28) Bardeen J, Cooper L N and Schrieffer J R Phys. Rev. **106** 162 (1957);

Theory of Superconductor J.R Scrieffer Advanced Boom Program United States of America (1999); D. M. Newns, P. C. Pattnaik, C. C. Tsuei, and C. C. Chi Phys. Rev. B **49** 3520 (1994)

**29)** H. Takagi, R. J. Cava, M. Marezio, B. Batlogg, J. J. Krajewski, W. F. Peck, Jr, P. Bordet, and D. E. Cox Phys. Rev Lett **68** 25 (1992); B.Nachumi, A.Keren and K.Kojima Phys. Rev. Lett. **77** 27 (1996)

30) A.Damascelli Review of Modern Physics **75** (2003)

31) N. E. Hussey, M. Abdel-Jawad, A. Carrington, A. P. Mackenzie and L. Balicas Nature **425** 814- 817 (2003)

32) J.P.Carbotte Reviews of Modern Physics **62** 4 (1990)

33) A. F. Bangura, P. M. C. Rourke, T. M. Benseman, M. Matusiak, J. R. Cooper, N. E. Hussey, and A. Carrington Phys. Rev. B **82**, 140501(R) (2010); Cyril Proust, Etienne Boaknin, R. W. Hill, Louis Taillefer, and A. P. Mackenzie Phys. Rev. Lett. **89**, 147003 (2002)

A. F. Santander-Syro, R. P. S. M. Lobo, N. Bontemps, Z. Konstantinovic, Z. Z. Li and H. Raffy EPL (Europhysics Letters), **62**, 4 (2003)